\documentclass[aps,twocolumn]{revtex4}
\usepackage {graphicx}
\usepackage{amsmath}
\usepackage{amssymb}

\newcommand{\mc}[1]{\mathcal{#1}}
\renewcommand{\vec}[1]{\boldsymbol{#1}}
\begin{document}

\title{Drift Lagrangian for relativistic particle in intense laser field}

\author{I.Y. Dodin$^{\dagger\,\ddagger}$ and N.J. Fisch}
\affiliation{$^{\dagger}$Princeton Plasma Physics Laboratory, Princeton, NJ 08543} 

\author{G.M. Fraiman}
\affiliation{$^{\ddagger}$Institute of Applied Physics RAS, 46 Ulyanova Str., Nizhnii Novgorod, Russia 603600} 

\begin{abstract}
The Lagrangian and Hamiltonian functions describing average motion of a relativistic particle under the action of a slightly inhomogeneous intense laser field are obtained. In weak low-frequency background fields, such a particle on average drifts with an effective, relativistically invariant mass, which depends on the laser intensity. The essence of the proposed ponderomotive formulation is presented in a physically intuitive and mathematically simple form, yet represents a powerful tool for studying various nonlinear phenomena caused by interaction of currently available smooth ultra-intense laser pulses with plasmas.\\[5pt] PACS numbers: 52.35.Mw, 52.20.Dq, 52.27.Ny, 45.20.Jj

\end{abstract}

\maketitle

The latest advances in high-power laser technology have resulted in the development of laser systems capable of delivering superstrong electromagnetic pulses, which can be focused to intensities as high as $10^{21}$ W/cm$^2$ \cite{RefPerry}, with even more powerful systems coming up in the near future \cite{RefMalkin}. The currently obtainable laser fields can accelerate electrons up to ultra-relativistic oscillatory velocities, previously unachievable in experiments on laser-matter interaction. This revolutionary progress is now giving a new life to theoretical studies on particle behavior under the action of intense electromagnetic radiation. The conventional models describing various nonlinear phenomena in plasmas illuminated by high-frequency radiation nowadays need to be revised, as new, ultra-powerful laser systems are becoming available for laboratory experiments. To explain the already observed phenomena and predict new effects taken place under the action of intense laser drive, adequate description of single-particle motion under relativistically strong radiation must be developed first.

Currently, particle motion is well-understood when the only forces present are those from the wave of uniform intensity \cite{RefLandau2}. However, to study the guiding center dynamics in inhomogeneous laser radiation or drifts determined by the presence of low-frequency background fields, additional analysis is needed. Expanding the conventional understanding to this area would result in a substantial progress in studying a number of plasma physics problems, such as, e.~g., Coulomb collisions and energetic particle production in strong laser fields. Moreover, the hydrodynamics and the electrodynamics of laser-illuminated plasmas would readily be available for general revision.

Often, the dynamics of a particle moving in a high-frequency field is described in terms of the ponderomotive approach. In nonrelativistic ponderomotive description, the effect of high-frequency electromagnetic forces on a particle is replaced by particle interaction with an average potential, linear on the intensity of laser radiation \cite{RefMiller2}. When ultra-intense lasers are employed, this conventional description needs to be generalized to relativistic motion. Contrary to the degenerate case of a circularly polarized field, in which high-frequency variations of the relativistic mass can be neglected \cite{RefLitvak}, the problem of particle motion in the case of a linear or elliptic polarization represents a certain challenge, but still can be studied analytically.

To describe the drift particle dynamics in such fields, recently, multiple studies were performed. Under various approximations, it was shown that the oscillating particle guiding center drifts in a smooth laser field with an effective mass, which depends on the electromagnetic field intensity \cite{RefMora, RefKibble, RefMoore, RefBauer}. (In strongly nonuniform laser fields though, the particle dynamics is more complicated \cite{RefMora, RefNonad}.) The Hamiltonian treatment of the relativistic drift under intense laser drive has been proposed in Ref. \cite{RefTokman}, though the problem of interaction with low-frequency background fields has not been studied. The first steps towards developing the general formalism with such interaction were made in Ref. \cite{RefMoore}. However, only smooth (compared to the amplitude of oscillations) low-frequency background fields were taken into consideration and the relativistic drift motion equations were induced without proper justification. 

These shortcomings are overcome in our paper, which major emphasis is twofold. First, we propose a general, fully relativistic Lagrangian formulation of ponderomotive description of particle motion under the action of a quasi-monochromatic slightly inhomogeneous laser field. The proposed approach is physically intuitive yet more systematic and simple in comparison with those discussed previously. After natural generalization, it allows including particle interaction with weak background forces, additional to those from the laser field. The discussion on that aspect of the guiding center motion constitutes the second emphasis of our work. We show the effective mass concept to be applicable to ponderomotive description of relativistic particle motion in low-frequency background fields, including even ones of small spatial scale compared to the amplitude of oscillations. In the end, we discuss the most promising applications of the proposed formulation and summarize our main ideas. 
To start, consider particle motion under the action of a plane laser wave propagating in vacuum, with vector potential given by
\begin{eqnarray}
\vec{A}(\vec{r},t)=(mc^2/e)\,\vec{a}(\eta), \label{EqVectorA}
\end{eqnarray}
where $\eta=\omega t - \vec{k} \cdot \vec{r}$ stands for the phase of the wave, $\omega$ is the wave frequency, and $\vec{k}=\vec{z}^0 \omega/c$ represents the wavevector. The polarization of the wave will be assumed fixed though arbitrary. The magnitude of $\vec{a}$, $a=eE/mc\omega$ (where $E$ is the laser electric field), can be understood as the ratio of the momentum imparted by the wave field in a single oscillation to $mc$, meaning that relativistic effects become important at $a \gtrsim 1$. (For the wavelength $\lambda=2\pi c/\omega=1\;\mu$m, the intensity corresponding to $a\sim 1$ for electrons is about $10^{18}$~W/cm$^2$.)

In a certain, unique, frame of reference, in such a field the particle undergoes stationary oscillatory ``figure-eight'' motion in a linearly polarized wave or circular motion in a wave with circular polarization \cite{RefLandau2}. Averaging over the oscillations, one comes to the concept of the guiding center motion, which we study below. First, let us consider the variational principle that states the minimum value of the action
\begin{equation}
S= \int^{\,t_2}_{t_1} L\, dt, \label{EqDeltaS}
\end{equation}
where $L$ is the Lagrangian function of the particle motion to be realized on the true trajectory. On time scales $t_2-t_1$ large compared to the oscillation period, the major contribution to the action $S$ (linear on $t_2-t_1$) is provided by the time-averaged part of the Lagrangian, $\langle L \rangle$, while the contribution of the oscillatory Lagrangian into the integral (\ref{EqDeltaS}) remains small. Thus, the action $S$ is approximately given by $S=\int^{t_2}_{t_1} \langle L \rangle dt$, from where it follows that $ \langle L \rangle$ can be treated as the Lagrangian of the average, guiding center motion.

To obtain the form of the drift Lagrangian $ L_0 \equiv \langle L \rangle$, let us consider the latter in the frame of reference where the guiding center rests. In the new frame, the guiding center Lagrangian $\mc{L}_0$ can be nothing but a constant, which we put in the form  
\begin{equation}
\mc{L}_0=-m_{\rm eff}c^2,
\end{equation}
in analogy with the Lagrangian of a true particle with zero velocity. The formally introduced quantity $m_{\rm eff}$ playing a role of a new, effective mass is yet to be defined. The action (\ref{EqDeltaS}) is relativistically invariant and can be written as $S= \int^{\tau_2}_{\tau_1} \mc{L}_0\, d\tau$, where the time $\tau$ represents the proper time of the guiding center. Since $d\tau$ is invariant by definition (and thus, so is the Lagrangian $\mc{L}_0$), the quantity $m_{\rm eff}$ must also be relativistically invariant. Using
\begin{equation}
d\tau=dt\,\sqrt{1-\upsilon_0^2/c^2}, \qquad \mc{L}_0\,d\tau= L_0  \,dt,
\end{equation}
where $\vec{\upsilon}_0$ is the velocity of the guiding center in the original frame of reference, one gets the Lagrangian of the guiding center motion 
\begin{equation}
L_0=-m_{\rm eff}c^2 \sqrt{1-\upsilon_0^2/c^2}, \label{EqGCLagr}
\end{equation}
which formally coincides with the Lagrangian of a relativistic particle with mass $m_{\rm eff}$ moving with velocity $\vec{\upsilon}_0$. Since the original frame was chosen arbitrarily, the above expression represents the general form of $L_0$, where $m_{\rm eff}$ is left to be expressed in terms of the parameters of the laser field.

Let us calculate $L_0$ in a laboratory frame of reference where the particle has a nonzero average velocity $\vec{\upsilon}_0$. Instructive by itself, the derivation to follow will also provide us with a number of useful relations connecting the parameters of the particle drift and those related to the actual motion. To proceed, consider the Lagrangian of the particle true motion given by
\begin{equation}
L=-mc^2\sqrt{1-\frac{\upsilon^2}{c^2}}+\frac{e}{c}\left(\vec{\upsilon}\cdot \vec{A}(\eta)\right),\label{EqLagr} 
\end{equation}
which is a known periodic function of the phase $\eta$ rather than time $t$. Thus, in order to average $L$ over time, one needs to derive a relation connecting time averaged and phase averaged quantities. For an arbitrary quantity $f$, its time and phase averaged values given by
\begin{equation}
\langle f \rangle =\frac{1}{\Delta }\int \limits^{t+\Delta }_t f\, dt',\qquad \overline{f}=\frac{1}{2\pi} \int\limits_\eta^{\eta+2\pi}f\,d\eta',
\label{EqAvrs}
\end{equation}
where the limits of integration over the phase correspond to the limits of integration over the time (i.~e. $\eta=\eta(t)$), and the time interval $\Delta $ is defined as one on which the total phase change equals $2\pi$:
\begin{equation}
\Delta =\int \limits^{\eta+2\pi}_{\eta} \frac{dt}{d\eta}\,d\eta.\label{EqT}
\end{equation}
The time interval $\Delta $ coincides with the wave period $2\pi/\omega$ only if particle motion is nonrelativistic. However, generally, the phase time-derivative is given by
\begin{equation}
\frac{d\eta}{dt}=\omega\left(1-\frac{\upsilon_z}{c}\right)=\omega\left(\frac{\gamma-p_z/mc}{\gamma}\right),\label{EqDEtaDt}
\end{equation}
where $\gamma=\left(1-\upsilon^2/c^2 \right)^{-1/2}$ is the normalized relativistic energy ($\mc{E}= m\gamma c^2$) and $\vec{p}=m\gamma \vec{\upsilon}$ it the particle kinetic momentum. 

Since the original Lagrangian depends on $\eta$ (that is, on the combination $z-ct$, rather than $z$ and $t$ separately), there exists an invariant of motion given by
\begin{equation}
u \equiv \gamma-p_z/mc={\rm const}.\label{EqUConst}
\end{equation}
Substituting the above expressions into Eq. (\ref{EqAvrs}), one gets
\begin{equation}
\langle f \rangle =\overline{\gamma f}/\overline{\gamma}.\label{EqAvr}
\end{equation}
Note that the obtained formula is valid only in case when the electromagnetic wave (\ref{EqVectorA}) is propagating in vacuum. If the refraction index of the medium differs from unit, Eqs. (\ref{EqDEtaDt}) and (\ref{EqUConst}) need to be modified, and the relation between time- and phase-averaged quantities becomes more complicated \cite{RefTokman}.

From Eqs. (\ref{EqUConst}) and (\ref{EqAvr}), it follows that
\begin{equation}
\overline{\gamma}=\sqrt{1+(\overline{p}/mc)^2+\overline{a^2}}, \qquad \overline{\gamma}=\gamma_0\sqrt{1+\overline{a^2}},\label{EqGamma}
\end{equation}
where $\gamma_0=(1-\upsilon_0^2/c^2)^{-1/2}$, and 
\begin{equation}
\vec{\upsilon}_0 \equiv \langle \vec{\upsilon} \rangle = \overline{\vec{p}}/m\overline{\gamma}
\end{equation}
is the drift velocity of the particle (compare with the inexact expression given in Ref. \cite{RefMora}). Thus, $L_0$ can be put in the form (\ref{EqGCLagr}) with $m_{\rm eff}$ given by
\begin{equation}
m_{\rm eff}=m\sqrt{1+e^2\overline{A^2}/m^2 c^4}.\label{EqMassEff}
\end{equation}

The guiding-center Lagrangian (\ref{EqGCLagr}) with the expression (\ref{EqMassEff}) for the effective mass was also obtained in Ref. \cite{RefBauer} by a somewhat similar yet a complicated and not a straightforward procedure. In the cited work, Eq. (\ref{EqMassEff}) was supposed valid only in the frame of reference where $\vec{\upsilon}_0=0$. In fact, as shown above, it remains applicable for arbitrary $\vec{\upsilon}_0$, and, more than that, the actual value of $m_{\rm eff}$ must be relativistically invariant. To express the effective mass in the invariant form, let us notice that, in the laboratory frame where we chose the electric potential $\phi=0$ (see Eq. (\ref{EqLagr})), $\sqrt{A^2}$ coincides with the norm of the 4-vector potential $\sqrt{A_\alpha A^\alpha}$, $A^\alpha=(\phi,\vec{A})$. The latter is Lorentz-invariant \cite{RefLandau2}, and remains such after being averaged over relativistically invariant phase $\eta$. Thus, the expression for $m_{\rm eff}$, invariant to relativistic transformations, can be put in the following form:
\begin{equation}
 m_{\rm eff}=m\sqrt{1+\frac{e^2}{m^2 c^4}\;\overline{(A_\alpha A^\alpha)}}.\label{EqMassEffInv}
\end{equation}
Eq. (\ref{EqMassEffInv}) was also given in Refs. \cite{RefMoore} where the average particle motion was studied otherwise.

Reverting to the formula for the drift Lagrangian (\ref{EqGCLagr}) with the effective mass given by (\ref{EqMassEff}), the canonical momentum of the guiding center motion $\vec{P}_0$ equals the phase-averaged kinetic momentum $\overline{\vec{p}}=m_{\rm eff} \gamma_0 \vec{\upsilon}_0$, and thus the Hamiltonian function of the guiding center motion can be put in the form
\begin{equation}
H_0=\sqrt{m_{\rm eff}^2c^4+P_0^2 c^2}.\label{EqHamilton}
\end{equation}
Here $m_{\rm eff}$ may smoothly depend on the guiding center location $\vec{R}_0$ and time $t$ if the wave envelope is slightly nonuniform or time-dependent. Precisely, that means that the laser intensity ``seen'' by the particle changes insignificantly on one period of particle oscillations, so that the averaging (\ref{EqAvrs}) still makes sense, i~.e.
\begin{equation}
l \gg r_\sim,\qquad T \gg \Delta, \qquad l/\upsilon_0 \gg \Delta, \label{EqLT}
\end{equation}
where $l$ and $T$ are the spatial and the temporal scales of the wave envelope (for detailed analysis, see Refs. \cite{RefMora, RefNonad, RefDodin}).

An alternative derivation of Eq. (\ref{EqHamilton}) can be found in Ref. \cite{RefTokman}, where a sequence of canonical transformations of the original motion equations was shown to lead to a similar result. In the present paper, we showed this tedious procedure to be unnecessary for obtaining the expression for the drift Hamiltonian. Comparing to the cited work, the distinguishing advantage of the formulation proposed in the present paper is that, because of its apparent mathematical simplicity, this formulation allows easy generalization of the drift Lagrangian and Hamiltonian formalism on the case when the oscillating particles undergo weak acceleration by large-scale low-frequency forces satisfying (\ref{EqLT}). Interaction with these forces enters the expression for $L_0$ {\it additively} and, what is most important, can still be considered in the framework of the effective mass concept.

To show this, consider an oscillating relativistic particle interacting with a field governed by the 4-vector potential $A^\alpha_{\rm bg}=(\phi_{\rm bg},\vec{A}_{\rm bg})$, where the subindex ``bg'' stands for a background field, additional to the one of the laser wave. Assume that the field is weak:
\begin{equation}
eE_{\rm bg}/\gamma_0m_{\rm eff} c \ll \omega, \qquad eB_{\rm bg}/\gamma_0m_{\rm eff} c \ll \omega, \label{EqSmall}
\end{equation}
where $\vec{E}_{\rm bg}$ and $\vec{B}_{\rm bg}$ are the corresponding electric and magnetic fields. In this case, the background fields do not impact the oscillatory motion significantly. Thus, averaging of the kinetic term  $mc^2/\gamma$ in the Lagrangian leads to the same expression as in Eq. (\ref{EqGCLagr}) with $m_{\rm eff}$ given by Eq. (\ref{EqMassEffInv}). In the zeroth-order approximation with respect to the small parameters (\ref{EqSmall}), the average part of the Lagrangian corresponding to particle interaction with the background field can be expressed in terms of the quantity $A^\alpha_0=(\phi_0,\vec{A}_0)$ given by
\begin{equation}
 A^\alpha_0=\Big \langle A^\alpha_{\rm bg}\left(\vec{R}_0+\vec{r}_\sim\right) \Big\rangle. \label{RefU}
\end{equation}
The time-averaging procedure is invariant with respect to changing the drift frame of reference, i.~e. does not alter the Lorentz transformation properties of the quantity being averaged. Thus, $A^\alpha_0$ represents a true 4-vector and can be considered as a new, effective electromagnetic field. In terms of this field's potentials, the drift Lagrangian can be put in the following form:
\begin{equation}
L_0=-m_{\rm eff}c^2\sqrt{1-\frac{\upsilon^2_0}{c^2}}+\frac{e}{c}\left(\vec{\upsilon}_0\cdot \vec{A}_0\right)-e\phi_0.
\end{equation}

In certain applications, it is of interest to consider particle interaction with background fields having spatial scale $l_{\rm bg} \lesssim r_\sim$. If the drift velocity is small, so that the drift displacement on a single period $\upsilon_0 \Delta$ is small compared to $l_{\rm bg}$, the ponderomotive description still can be applied. However, in this case the difference between the time-averaged potential $A^\alpha_0$ and the true potential $A^\alpha_{\rm bg}$ taken at the location of the guiding center $\vec{R}_0$, is crucial. For example, this situation is realized at Coulomb scattering in intense laser fields when $r_\sim$ exceeds the radius of effective interaction \cite{RefMittleman}. Note that, as follows from the above analysis, the characteristic amplitude of the effective potential remains unchanged as one generalizes the expression for $\phi_0$ to the case of relativistic particle motion. In this case, the only difference in calculating $\phi_0$ is provided through the change in the oscillatory trajectory $\vec{r}_\sim(t)$ to be averaged over.

In the context of the Coulomb scattering problem, the considered Lagrangian approach represents a unique tool for studying ponderomotive and even stochastic behavior of particles being scattered. This problem deserves detailed consideration and will be discussed in future works, though, briefly, the extension of the proposed formulation can be explained as follows. Stochastic behavior of a dynamical system with periodic coefficients is often convenient to describe in terms of mapping of the dynamical trajectory onto a subspace of the system phase space (for review, see Ref. \cite{RefLichtenberg}). For the Hamiltonian mapping $(\vec{R}_0,\vec{P}_0) \rightarrow (\overline{\vec{R}}_0,\overline{\vec{P}}_0)$ connecting the particle locations and momenta before and after the time interval equal to the period of the laser field, the generating function is given by the action (\ref{EqDeltaS}) with $t_1=t$ and $t_2=t+2\pi/\omega$ \cite{RefLichtenberg}. Since the obtained drift Lagrangian is approximately proportional to $S$, it can readily be used for constructing the actual form of this mapping. As will be shown in our future publications, when studying the statistical properties of particle stochastic dynamics (rather than single particle motion) by means of such a mapping, the conditions (\ref{EqLT}) can be significantly relaxed, which substantially broadens the proposed Lagrangian approach applicability. That also allows a significant progress in studying the problem of energetic particle production in strong laser fields \cite{RefEnergetic}.

Since, in the case of relativistic drift, $\vec{r}_\sim$ depends on $\vec{\upsilon}_0$, the expression for the canonical momentum $\vec{P}_0=\partial L_0/\partial \vec{\upsilon}_0$ the drift motion equations become complicated. However, in two special cases of interest, those can be simplified. In a large-scale background field satisfying the conditions (\ref{EqLT}), locally, $A^\alpha_{\rm bg}$ can be treated as a linear function of $\vec{r}$. Therefore, the velocity-dependent part averages out when calculating the potential $A^\alpha_0$, and one gets $A^\alpha_0\approx A^\alpha_{\rm bg}$. Thus, the drift canonical momentum equals $\vec{P}_0=m_{\rm eff} \gamma_0 \vec{\upsilon}_0+(e/c)\vec{A}_{\rm bg}$, and the Hamiltonian function is given by
\begin{equation}
H_0=\sqrt{m_{\rm eff}^2c^4+\left(\vec{P}_0-\frac{e}{c}\vec{A}_{\rm bg}\right)^2 c^2}+e\phi_{\rm bg},\label{EqHamBG}
\end{equation}
where the potentials are assumed to be slow functions of $\vec{R}_0$ and $t$. The guiding center motion equations can be put in the covariant form
\begin{equation}
\frac{dR^\alpha_0}{d\tau}=\frac{p^\alpha_0}{m_{\rm eff}},\qquad \frac{dp^\alpha_0}{d\tau}=\frac{e}{c}F^{\alpha\beta}_{\rm bg}U_\beta-c^2\frac{\partial m_{\rm eff}}{\partial R^\alpha_0},\label{EqMotionEqs}
\end{equation}
where $R^\alpha_0=(ct,\vec{R}_0)$ is the 4-coordinate of the guiding center, $p^\alpha_0=(\mc{E}_0/c,m_{\rm eff}\gamma_0\vec{\upsilon}_0)$ is the drift kinetic 4-momentum, $\mc{E}_0=m_{\rm eff}\gamma_0 c^2$ is the energy of the guiding center motion, $F^{\alpha\beta}_{\rm bg}$ is the electromagnetic field tensor corresponding to the potential $A^\alpha_{\rm bg}$ \cite{RefLandau2}, and $U^\alpha=\gamma_0(c,\vec{\upsilon}_0)$ is the guiding center 4-velocity. Covariant Eqs. (\ref{EqMotionEqs}) were also given in Ref. \cite{RefMoore}, though no strict derivation of those was proposed. Another expression for the relativistic ponderomotive force is given in Ref. \cite{RefBauer}.

From Eq. (\ref{EqHamBG}), it follows that, in low-frequency large-scale background field, the guiding center of a relativistic particle moving under the action of intense laser radiation behaves as a particle with the effective mass $m_{\rm eff}$ drifting in the same background field. For example (and in coincidence with the results obtained in Ref. \cite{RefMoore}), in static magnetic field $\vec{B}_{\rm bg}$, the guiding center undergoes Larmor motion with the cyclotron frequency $\omega_B=eB_{\rm bg}/\gamma_0m_{\rm eff}c$. Conventional expression for the drift velocity in nonuniform magnetic field \cite{RefLandau2} also readily applies to the average motion if the particle true mass is replaced with the one given by Eq. (\ref{EqMassEffInv}).

In addition to the case of large-scale background fields, the guiding center motion equations can also be put in a simple, physically intuitive form in the case of nonrelativistic drift motion. Since the drift velocity enters the expression for $A^\alpha_0$ only through relativistic dependence of $\vec{r}_\sim$ on $\vec{\upsilon}_0/c$, then, in the case $\vec{\upsilon}_0 \ll c$, $\partial A^\alpha_0/\partial \vec{\upsilon}_0$ can be neglected. In this case, the drift canonical momentum is given by $\vec{P}_0=m_{\rm eff} \vec{\upsilon}_0+(e/c)\vec{A}_0$, and the Hamiltonian can be put in the form
\begin{eqnarray}
H_0=\frac{1}{2m_{\rm eff}}\left(\vec{P}_0-\frac{e}{c}\vec{A}_0\right)^2+m_{\rm eff} c^2+e\phi_0,\label{EqNonrelHam}
\end{eqnarray}
where the effective mass $m_{\rm eff}$ and the potential energy $\psi_{\rm eff}=m_{\rm eff}c^2+e\phi_0$ may slowly depend on the guiding center location $\vec{R}_0$ and time $t$. Note that even in uniform laser field, $\nabla \psi_{\rm eff}$ may differ significantly from $e\nabla \phi_{\rm bg}$ when the amplitude of particle oscillations $r_\sim$ exceeds the spatial scale of the background field $l$ \cite{RefMittleman}. The regime of slow drift motion superimposed on relativistic oscillations is the one, which is actually realized in many current experiments on intensive laser pulses interaction with rare plasmas. (Under rare plasmas we assume those having a refraction index close to unity, as assumed for all the results obtained in the present paper.) This fact makes the above analysis especially useful from the practical point of view, as it represent a simple tool for studying the actual experimental data. Finally, the well-known nonrelativistic ponderomotive potential \cite{RefMiller2} can be readily derived from Eq. (\ref{EqNonrelHam}) by keeping the correction to the effective mass, linear with respect to the wave intensity (see also Ref. \cite{RefBauer}).

In summary, we showed that, in weak low-frequency background fields, relativistic particle moving under the action of intense laser radiation drifts like a quasi-particle with an effective mass, which depends on the intensity of the laser field. The intuitive expectation that, by the order of magnitude, the drift motion equations must coincide with those without the laser field if the appropriate relativistic correction of particle mass is introduced, can now be considered proven for various types of background fields. The proposed formulation can be useful for studying numerous phenomena resulting from intense laser-plasma interaction, such as, e.~g., the energetic particle production and Coulomb scattering in strong laser fields. Moreover, the mathematical simplicity of the proposed approach allows easy generalization of the rare plasma hydrodynamics and electrodynamics to the case of plasmas illuminated by ultra-intense laser radiation. Replacing the electron mass with the effective mass (\ref{EqMassEffInv}), one can readily derive the generalized dispersion relations for various linear waves in plasmas, as well as revise the nonlinear plasma dynamics.

The work was supported by the US DOE, under contract DE-AC02-76 CHO3073, and Russian Foundation for Basic Research, grants 02-02-17277, 02-02-17275.

\end{document}